\documentclass[10.5pt,onecolumn]{article}
\usepackage[utf8]{inputenc}
\usepackage[english]{babel}
\usepackage[numbers]{natbib}
\usepackage[T1]{fontenc}
\usepackage[margin=2.5cm]{geometry}
\usepackage{graphicx}%
\usepackage{xurl}
\usepackage[ruled]{algorithm2e}
\usepackage{algpseudocode}
\graphicspath{{./figures/}}
\usepackage{multirow}%
\usepackage{amsmath,amssymb,amsfonts}%
\usepackage{amsthm}%
\usepackage{mathrsfs}%
\usepackage[title]{appendix}%
\usepackage{xcolor}%
\usepackage{textcomp}%
\usepackage{manyfoot}%
\usepackage{booktabs}%
\usepackage{listings}%
\usepackage{tikz,lipsum,lmodern}
\usepackage{amsmath,amssymb,amsthm}

\theoremstyle{definition}

\usepackage{enumerate}
\usepackage{qcircuit}
\usepackage{todonotes}

\usepackage{caption}
\usepackage{subcaption}
\usepackage[most]{tcolorbox}
\usepackage{venndiagram}

\usepackage{hyperref}
\newcommand{\ket}[1]{\ensuremath{\left|#1\right\rangle}} 
\newcommand{\bra}[1]{\ensuremath{\left\langle#1\right|}} 
\newcommand{\braket}[2]{\ensuremath{\left\langle#1|#2\right\rangle}}

\newcommand{\norm}[1]{\ensuremath{\lVert#1\rVert }}
\newcommand{\revfix}[1]{{1}}
\let\vec\mathbf

\usepackage{parskip}
\usepackage{titling}
\usepackage{authblk}

\title{An alternative explicit circuit diagram for the quantum search algorithm by implementing a non-unitary gate}
\author{Ammar Daskin}
\affil{
Department of Computer Engineering\\
Istanbul Medeniyet University\\ 
Istanbul, Turkiye, 34000\\
Email Address: adaskin25@gmail.com\\
Orcid ID: \href{https://orcid.org/0000-0002-1497-5031}{0000-0002-1497-5031}
}
\date{}
\begin{document}

\maketitle

\begin{abstract}
Since the final quantum state in the Grover search algorithm is the normalized marked quantum state from the Gram-Schmidt process, Abrams and Lloyd \cite{abrams1998nonlinear} has shown that we can generate this vector by using a non-unitary gate.
Following their ideas, in this paper, we present multiple explicit unitary implementations by using the square root of the non-unitary matrix and by a unitary matrix that mimics the Gram-Schmidt process.  We also discuss the implementation through a linear combination of unitary matrices or similar methods and how these approximations may change the complexity. 
The reading of the marked element from the given circuits with high probability still requires multiple repetitions similar to the original algorithm. However, it gives an alternative implementations which may be useful in certain platforms. In addition, in the appendix of the paper, we show that the circuits can be used to group set elements which can be integrated into different algorithmic schemes.
\end{abstract}


\section{Introduction}
The Grover search algorithm\cite{grover1996fast,grover1997quantum} is an iterative algorithm consisting of two operators \cite{nielsen2010quantum}: $U_{mark}$ that marks (changes the sign of) the solution. The marking, for instance, can be done by controlling this operator by  a register that represents the superposition state of the output of a function.
The second operator is $U_{amplify}$ which increases the magnitude of the amplitude that is marked by the previous operator.
After applying these two operators $O(\sqrt{N})$ times for a state of dimension $N$, the marked solution can be measured with probability $\approx 1$.  The algorithm is generalized for searching multiple elements \cite{biron1999generalized} and through the years its many variants and implementations are presented \cite{giri2017review}.
We can also understand the algorithm through the Gram-Schmidt orthogonalization process \cite{golub2013matrix} used in linear algebra. In our case, the process start with assuming the initial state vector as $\mathbf{v_0}$, generally a superposition state,  and the marked state as $\mathbf{v_1}$. Here, these two vectors differ only on the sign of one element and they are not orthogonal to each other: in the case when $\mathbf{v_0}$ is a superposition state $\mathbf{v_0^Tv_1} =  \frac{N-2}{N}$.

Abrams and Lloyd \cite{abrams1998nonlinear} has shown that
by allowing non-linearities, e.g. using a non-unitary gate on an ancilla register, one can generate this solution state directly on quantum computers.   The circuit for this model is investigated in \cite{terashima2005nonunitary}. 
This model is also used to study different problems: e.g. studies related to general non-unitary computing \cite{gingrich2004non,williams2004probabilistic}, simulation of non-unitary matrices \cite{daskin2017ancilla, mazzola2019nonunitary}, exploring the divergence of fisher information \cite{waghela2024simulation}. 

In the following sections, we first review the key steps in the Gram-Schmidt process, then follow mainly Ref. \cite{abrams1998nonlinear} to discuss and formalize how to construct this process on quantum states which requires a non-unitary gate. We then implement this non-unitary gate as a unitary circuit by using additional ancillary qubits.  We also discuss how the implementation can done through the linear combination of unitaries.
We then show how the amplitude amplification can be used to amplify the success probability of the desired part.  After discussing how the approach can be used for multiple database search, we discuss its complexity. In the appendix, we define the steps and give explicit circuits to show how to group elements of a given set.

\section{The Gram-Schmidt orthogonalization process}
Gram-Schmidt process is used to orthogonalize given vectors. 
It is used to construct QR decomposition of a matrix. 
An alternative quantum version \cite{zhang2021quantum} is also proposed which is used to construct orthonormal basis vectors of a quantum state from a given arbitrary vectors.

In the Gram-Schmidt process\cite{golub2013matrix} the first vector does not change if it is normalized.
\begin{equation}
    \mathbf{u_0} = \mathbf{v_0}.
\end{equation}
The second vector is found by removing the residual in their inner product:
\begin{equation}
\label{Eq:u1}
    \mathbf{u_1} = \mathbf{v_1} - \alpha \mathbf{u_0}, \mathbf{\tilde{u}_1} = \frac{\mathbf{u_1}}{\norm{\mathbf{u_1}}},
\end{equation}
where $\mathbf{\tilde{u}_1}$ represents the normalized version of the vector $\mathbf{u_1}$.
For the normalized vectors;
\begin{equation}
    \alpha=\frac{\braket{\mathbf{v_1}}{\mathbf{u_0}}}{\braket{\mathbf{u_0}}{\mathbf{u_0}}}
    = \braket{\mathbf{v_1}}{\mathbf{v_0}} = \frac{N-2}{N}.
\end{equation}

As an example consider the following two vectors:
\begin{equation}
\mathbf{v_0} = \frac{1}{2} \left( \begin{matrix}
1\\ 1\\ 1\\ 1
    \end{matrix}
    \right), \mathbf{v_1} = \frac{1}{2}\left( \begin{matrix}
1\\ 1\\ -1\\ 1
    \end{matrix}
    \right).    
\end{equation}
The solution is chosen as the third element and marked by negative sign. 
Then we can define the  orthogonal vectors: 
$\mathbf{u_0} = \mathbf{v_0}$ and $\mathbf{u_1}$ as:
\begin{equation}
\begin{split}
\mathbf{u_1} = & \mathbf{v_1} - \frac{\braket{\mathbf{v_1}}{\mathbf{u_0}}}{\|\mathbf{u_0}\|}\mathbf{u_0} \\
= &  \frac{1}{2}\left( \begin{matrix}
1\\ 1\\ -1\\ 1
    \end{matrix}
    \right) - \frac{1}{4} \left( \begin{matrix}
1\\ 1\\ 1\\ 1
    \end{matrix}
    \right) =  \left( \begin{matrix}
1/4\\ 1/4\\ -3/4\\ 1/4
    \end{matrix}
    \right) .
\end{split}
\end{equation}
Therefore, the amplitude of the marked state after the orthogonalization becomes 3 times of the others. And the probability becomes three times of the sum of other probabilities.
Note that normalization of $\mathbf{v_0}$ and $\mathbf{v_1}$ does not change this fact since it scales all the elements.
In generic case with all elements one in $\mathbf{u_0}$, we can define the terms as:
\begin{equation}
\begin{split}
&\braket{\mathbf{v_1}}{\mathbf{u_0}} = N-2 \text{ if normalized }\braket{\mathbf{v_1}}{\mathbf{u_0}} = \frac{N-2}{N}\\
&\|\mathbf{u_0}\| = \sqrt{N}, \text{ if normalized } \|\mathbf{u_0}\| = 1
\end{split}
\end{equation}
Therefore, we can define the vector elements of $\mathbf{u_1}$ as:
\begin{align}
&\mathbf{u_1}_{unmarked} = \frac{1}{\sqrt{N}}-\frac{N-2}{N\sqrt{N}} = \frac{2}{N\sqrt{N}},\\
&\mathbf{u_1}_{marked} = -\frac{1}{\sqrt{N}}-\frac{N-2}{N\sqrt{N}} = \frac{2-2N}{N\sqrt{N}}
\end{align}
If we take the ratio of the amplitudes for the marked and unmarked elements given above,  we see that overall the marked element is $(N-1)$ times greater than the other elements.
This orthogonalization mirrors Grover’s amplitude amplification.
Therefore, we can consider this as the final state in the Grover search algorithm.

Note that in the Grover search algorithm, the final orthogonal vector is found through small changes by applying the $U_{mark}$ and $U_{amplify}$ operators successively. 

\subsection{Constructing $\mathbf{u_1}$ as a quantum state}
Since we know every vector and value except the index of the signed vector element (solution) in the above equations, 
we can simply try to build Eq.\eqref{Eq:u1} as a quantum state.

Therefore, our first main goal is to construct unnormalized $\mathbf{u_1}$ by using a non-unitary gate and ancilla register. This will allow us to form a state similar to $\mathbf{u_1}$ inside a larger system.

For this purpose, we will consider to first construct the following vector (normalized):
\begin{equation}
    \frac{1}{\sqrt{2}}\left( \begin{matrix}
        \mathbf{v_0}\\
        \mathbf{v_1}
    \end{matrix}
    \right).
\end{equation}
Suppose we have the circuit for the marking part and we use the equal superposition state for $\mathbf{v_0}$.

\begin{figure}[t!]
\begin{center}
        \Qcircuit @C=1em @R=.7em {
\ket{0}&&\qw & \gate{H} & \ctrl{1} & \qw& \qw \\
\ket{\mathbf{0}}&&/\qw & \gate{H} & \multigate{1}{U_{mark}} & \qw& \qw \\
\ket{\mathbf{0}}&&/\qw & \gate{H}& \ghost{U_{mark}} & \qw & \qw
}
\end{center}
\caption{\label{fig:partial}Partial circuit.}
\end{figure}
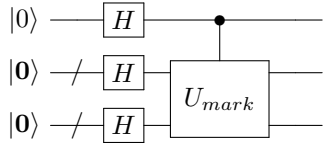

Abrams and Lloyd \cite{abrams1998nonlinear} has showed that,
we can generate the above state on quantum circuit by using a nonlinear quantum operator. We depict this  with a non-unitary gate in Fig.\ref{fig:partial}.

In terms of quantum states, after the Hadamard gates and the controlled $U_{mark}$, the circuit yields the following state:
\begin{equation}
\frac{1}{\sqrt{2}}\ket{0}\ket{\mathbf{v_0}}+\frac{1}{\sqrt{2}}\ket{1}\ket{\mathbf{v_1}}.
\end{equation}
If we apply another Hadamard gate to the first qubit, the states becomes:
\begin{align}
&\frac{1}{2}\left(\ket{0}+\ket{1}\right)\ket{\mathbf{v_0}}+\frac{1}{2}\left(\ket{0}-\ket{1}\right)\ket{\mathbf{v_1}}\\
&= \frac{1}{2}\ket{0}\left(\ket{\mathbf{v_0}}+\ket{\mathbf{v_1}}\right)+\frac{1}{2}\ket{1}\left(\ket{\mathbf{v_0}}-\ket{\mathbf{v_1}}\right).
\end{align}
We can write \ket{\mathbf{v_0}} and \ket{\mathbf{v_1}} as:
\begin{equation}
\ket{\mathbf{v_1}} = \ket{\mathbf{rest}} + \ket{x} \text{ and } \ket{\mathbf{v_1}} = \ket{\mathbf{rest}} - \ket{x},
\end{equation}
where $x$ is the index of the marked element and  \ket{\mathbf{rest}} represents the remaining parts of the vector (the place represented by $x$ is a zero element in the \ket{\mathbf{rest}} and only non-zero element in \ket{x}). Note that these vectors alone are not normalized and similar to ``good" and ``bad" states in quantum counting \cite{brassard1998quantum}.
Then if we substitute this with $\mathbf{v_1}$ and $\mathbf{v_0}$, 
the state turns into the following simple form:
\begin{equation}
 \ket{0}\ket{\mathbf{rest}}-\ket{1}\ket{x}.
\end{equation}
The above equation just makes the solution as conditioned on the first qubit state. 
However, it does not change its probability which is still $1/N$.
\subsection{Using a non-unitary operator}
Instead of the last Hadamard gate, we can use a matrix which does not have uniformly scaled elements.
Assume we are able to use a quantum gate described by the following matrix with generic real values:
\begin{equation}
  M =   \left(\begin{matrix}
        a&b\\
        c&d
    \end{matrix}
    \right).
\end{equation}
Rewriting the equations where the Hadamard gate is used, we obtain the following: 
\begin{equation}
\begin{split}
&\frac{1}{\sqrt{2}}\left(a\ket{0}+b\ket{1}\right)\ket{\mathbf{v_0}}+\frac{1}{\sqrt{2}}\left(c\ket{0}+d\ket{1}\right)\ket{\mathbf{v_1}}\\
&= \frac{1}{\sqrt{2}}\ket{0}\left(a\ket{\mathbf{v_0}}+c\ket{\mathbf{v_1}}\right)+\frac{1}{\sqrt{2}}\ket{1}\left(b\ket{\mathbf{v_0}}+d\ket{\mathbf{v_1}}\right).
\end{split}
\end{equation}
Similarly, if we write in terms of $\ket{x}$ and \ket{\mathbf{rest}}:
\begin{equation}
\begin{split}
\ket{\psi} = & \frac{1}{\sqrt{2}}
\ket{0}\left(a\ket{\mathbf{rest}} + a\ket{x}+c\ket{\mathbf{rest}} - c\ket{x}\right)
\\ &+\frac{1}{\sqrt{2}}\ket{1}\left(b\ket{\mathbf{rest}} + b\ket{x}+d\ket{\mathbf{rest}} + -d\ket{x}\right)\\
= & \frac{1}{\sqrt{2}}\ket{0}\left((a+c)\ket{\mathbf{rest}}+\left(a-c\right)\ket{x}\right)\\
&+ \frac{1}{\sqrt{2}}\ket{1}\left(\left(b+d\right)\ket{\mathbf{rest}}+\left(b-d\right)\ket{x}\right).
\end{split}
\end{equation}

\begin{figure}[t]
\begin{center}
        \Qcircuit @C=1em @R=.7em {
\ket{0}&&\qw & \gate{H} & \ctrl{1} & \qw& \gate{M}&\qw \\
\ket{\mathbf{0}}&&/\qw & \gate{H} & \multigate{1}{U_{mark}}& \qw&\qw& \qw \\
\ket{\mathbf{0}}&&/\qw & \gate{H}& \ghost{U_{mark}} & \qw& \qw & \qw
}
\end{center}
    \caption{Search algorithm circuit. $M$ is a non-unitary gate.}
    \label{fig:circuitM}
\end{figure}
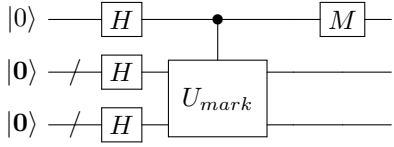

To maximize the probabilities of $\ket{x}$, as also done  by  Abrams and Lloyd \cite{abrams1998nonlinear}, we can conclude the matrix as:
\begin{equation}
M =        \left(\begin{matrix}
        a&b\\
        c&d
    \end{matrix}
    \right)=  \frac{1}{2}  \left(\begin{matrix}
        1&-1\\
        -1&1
    \end{matrix}
    \right).
\end{equation}
The resulting circuit is represented in Fig.\ref{fig:circuitM}.
In circuit, after the application of $M$ we obtain the following final quantum state:
\begin{equation}
\label{eq:psifinal}
\ket{\psi} =  \frac{1}{\sqrt{2}}\ket{0}\ket{x}+ \frac{1}{\sqrt{2}}\ket{1}\ket{x}.
\end{equation}
 Measurement on the output gives the solution $\ket{x}$.

\begin{figure}
     \centering
     \begin{subfigure}[t]{0.45\columnwidth}
     \includegraphics[width=1\linewidth]{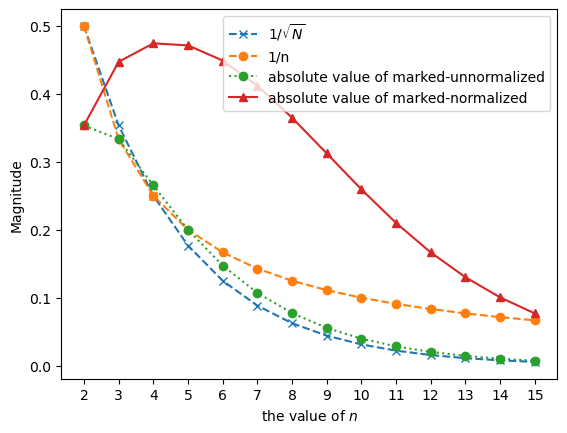}
     \caption{The change of the magnitude of the marked element if we use $(n-2)/n$ as in Eq.\eqref{eq:barm}. }
     \label{fig:magnitude of markedn-2}
 \end{subfigure}
 \begin{subfigure}[t]{0.45\columnwidth}
     \centering
     \includegraphics[width=1\linewidth]{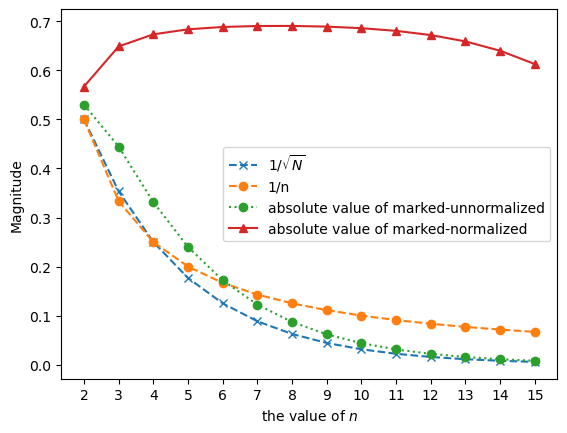}
     \caption{The change of the magnitude of the marked element if we use $(n^2-2)/n^2$  in Eq.\eqref{eq:barm}.  }
     \label{fig:magnitude of markednsquared}
 \end{subfigure}
 \caption{Change in the absolute value of the amplitude for the marked element for different number of qubits. Note that the overall unnormalized magnitude is twice the values reported in the graph since we have the marked state when the first qubit \ket{0} and we have another when  the first qubit is in \ket{1}.}
 \end{figure}

 Here, $M$ is a singular matrix with a zero eigenvalue. Instead of this we can insert the value of $\alpha$ into this which also makes the matrix non-singular with real eigenspectrum:
 \begin{equation}
     \tilde{M} =  \frac{1}{\rho}  \left(\begin{matrix}
        -\frac{N-2}{N}&1\\
        1&-1
    \end{matrix}
    \right) \text{ or } \frac{1}{\rho}  \left(\begin{matrix}
        -\frac{N-2}{N}&1\\
        1&-\frac{N-2}{N}
    \end{matrix}
    \right),
 \end{equation}
 where $\rho$ is a normalization constant. Note the matrix has one small and one big eigenvalue. And the resulting quantum state involves only two non-zero elements. Although $\tilde{M}$ is not singular, it is still exponentially close to one. Therefore, instead  of using $\frac{N-2}{N}$, we can try to use $\frac{poly(n)-2}{poly(n)}$. For instance; we can use:
 \begin{equation}
 \label{eq:barm}
          \bar{M} =  \frac{1}{\rho}  \left(\begin{matrix}
        -\frac{n-2}{n}&1\\
        1&-1
    \end{matrix}
    \right).
 \end{equation}
 Again $\rho$ is just a scaling, normalization, constant.
 Note that this would make much easier to implement the gate. However, this will introduce more non-zero elements to the output state even though those elements would have less magnitude.
 On randomly chosen marked element and for $\rho=1$, Fig.\ref{fig:magnitude of markedn-2} shows the change of its magnitude (the absolute value of the amplitude for the marked, $x$th, element in the state without applying any measurement on any qubits) for different number of qubits. 
 If we use the coefficient $(n^2-2)/n^2$, then the magnitude of the marked element further dominates the output as shown in Fig.\ref{fig:magnitude of markednsquared}.
 Also note that as in the Grover search algorithm, one can start with an $M$ which is easy to implement and iterate the application of this to the quantum state in combination with the $U_{mark}$.
 
\section{Unitary circuit implementations and the amplitude amplification}
Various unitary implementations for the non-unitary gates are proposed in literature for different purposes (e.g.\cite{zheng2021universal, xu2022nonlinear}). Also note that several similar gate implementations of thenonlinear model \cite{abrams1998nonlinear} are reported in \cite{terashima2005nonunitary}.

Here, we first consider the following matrices:
 \begin{equation}
     J =  \frac{1}{\kappa_J}  \left(\begin{matrix}
        1&1\\
        1&1
    \end{matrix}
    \right) \text{ and } M = \frac{1}{\kappa_M}  \left(\begin{matrix}
        a&-1\\
        -1&a
    \end{matrix}
    \right),
 \end{equation}
where for now $\kappa$s are scaling constants and we use the symbol $a$ again to represent a real number.

The eigenspaces of both matrices $M$ and $J$ are formed by the Hadamard matrix:
\begin{equation}
\label{eq:matrixJ}
\begin{split}
     J = &  H^T D_J H \\ = & \frac{1}{2}\left( \begin{matrix}
       1& 1\\
       -1 &1
    \end{matrix}\right)\left( \begin{matrix}
       0& 0\\
       0 &1
    \end{matrix}\right)
    \left( \begin{matrix}
       1& -1\\
       1 &1
    \end{matrix}\right) \\ = &  \frac{1}{2} \left( \begin{matrix}
       1& 1\\
       1 &1
    \end{matrix}\right)    
\end{split}
\end{equation}
Similarly,  $M$ has the eigendecomposition:
\begin{equation}
\begin{split}
    M = & H^T D_M H \\ = & \frac{1}{2}\left( \begin{matrix}
       1& 1\\
       -1 &1
    \end{matrix}\right)\left( \begin{matrix}
       a+1& 0\\
       0 &a-1
    \end{matrix}\right)
    \left( \begin{matrix}
       1& -1\\
       1 &1
    \end{matrix}\right) \\ =  & \frac{1}{2}\left( \begin{matrix}
       a+1& a-1\\
       -(a+1) &a-1
    \end{matrix}\right)
    \left( \begin{matrix}
       1& -1\\
       1 &1
    \end{matrix}\right)\\ = &\left( \begin{matrix}
       a& -1\\
       -1 &a
    \end{matrix}\right) 
\end{split}
\end{equation}

Now, consider as an example the following quantum operator on a $k-$qubit register:
\begin{equation}
\label{eq:JotimesM}
        C = J \otimes J\otimes \dots \otimes M. 
\end{equation}
Since the matrix we have so far used is the non-unitary, it can be implemented via block encoding. In our case, we will construct the unitary matrix by using the following the matrix with the square roots \cite{daskin2017ancilla}:
\begin{equation}
    \mathfrak{C} = \left( \begin{matrix}
        \frac{C}{\kappa_C}& -\sqrt{1- \frac{C^2}{\kappa_C^2}}\\
       \sqrt{1- \frac{C^2}{\kappa_C^2} }& \frac{C}{\kappa_C}
    \end{matrix}\right) = \left( \begin{matrix}
        \hat{C}& -S\\
       S & \hat{C}
    \end{matrix}\right).
\end{equation}
If the eigenvalues of $C$ determined via the matrices $M$ and $J$ and so by the value of $a$ are in the range [0, 1], 
the above square roots are defined in the real space and $S=S^T$.
(The eigenvalues can be scaled into the region [0,1] by using $\kappa_C = (a+1)$.) 
After this scaling and finding the square matrix,  $\mathfrak{C}$ can be used as a unitary matrix.
Note that if we use $k=1$, $C$ becomes the matrix $M$.

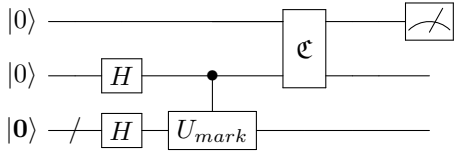
\begin{figure}
\begin{center}
        \Qcircuit @C=1em @R=.7em {
\ket{0}&&\qw &  \qw & \qw &\multigate{1}{\mathfrak{C}}\qw& \qw & \qw& \meter \\
\ket{0}&&\qw & \gate{H} & \ctrl{1}&\ghost{\mathfrak{C}} &\qw & \qw& \qw&\\
\ket{\mathbf{0}}&&/\qw & \gate{H} & \gate{U_{mark}} 
& \qw&\qw& \qw&\qw  \\
}
\end{center}
    \caption{ Unitary search algorithm circuit. If the measurement output on the first qubit is \ket{0}, we have the solution in the last qubits. }
    \label{fig:circuitUnitaryC}
\end{figure}

The circuit for this unitary can be easily drawn because its eigendecomposition is known. The circuit applies the scaled matrix $C$, when a new ancilla qubit is in \ket{0} state. This circuit can be depicted as in Fig.\ref{fig:circuitUnitaryC}. 

\begin{figure*}[t!]
    \centering
    \begin{subfigure}[t]{0.45\columnwidth}
        \centering
        \includegraphics[width=1\columnwidth]{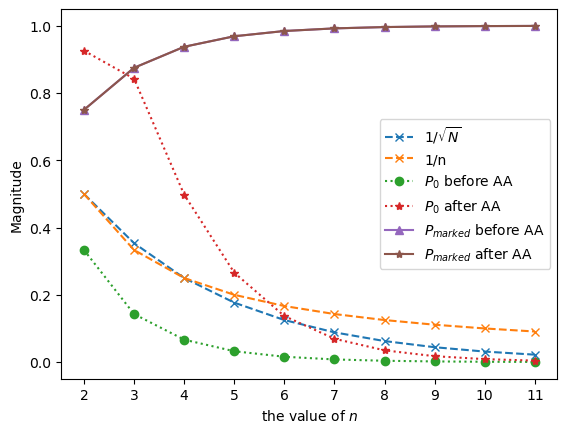}
        \caption{$a = \frac{N-2}{N}$.}
    \end{subfigure}
    \begin{subfigure}[t]{0.45\columnwidth}
        \centering
        \includegraphics[width=1\columnwidth]{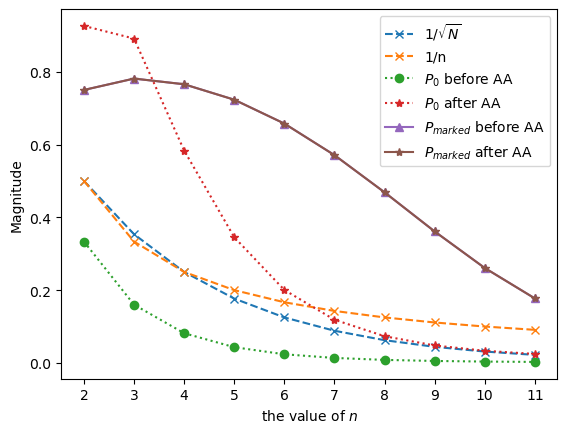}
        \caption{$a = \frac{n-1}{n}$ }
    \end{subfigure}\\
        \begin{subfigure}[t]{0.45\columnwidth}
        \centering
        \includegraphics[width=1\columnwidth]{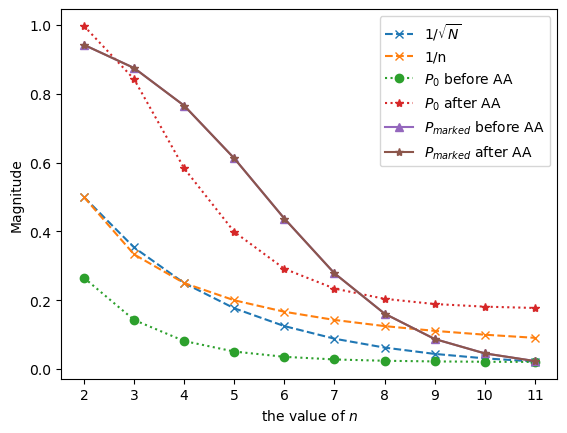}
        \caption{$a = 0.75$.}
    \end{subfigure}
    \begin{subfigure}[t]{0.45\columnwidth}
        \centering
        \includegraphics[width=1\columnwidth]{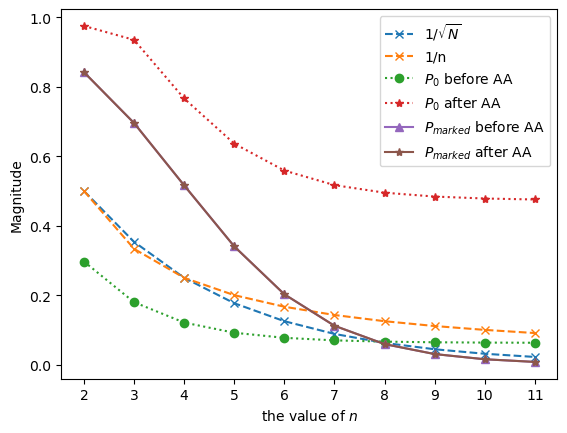}
        \caption{$a = 0.6$. }
    \end{subfigure}
       \caption{\label{fig:simulationofC}
       For different $a$ values in matrix $M$ simulation of the circuit with the unitary matrix $\mathfrak{C}$ and \textbf{1-step} of the  amplitude amplification.}
    \end{figure*}

We have simulated this circuit by using different values for $a$ in the matrix $M$. In particular, if we use $a$ close to 1, this increase the success probability of the marked state in the collapsed state. However, the probability to see the first qubit in \ket{0} state diminishes.
This is shown in Fig.\ref{fig:simulationofC}.
\subsection{An alternative unitary circuit by simulating Gram-Schmidt process}
An alternative circuit can be formed by using the following matrix:
\begin{equation}
    \left(\begin{matrix}
        \ket{\psi}\bra{\psi} & I-\ket{\psi}\bra{\psi}\\
        I-\ket{\psi}\bra{\psi} &  \ket{\psi}\bra{\psi}
    \end{matrix}\right), 
\end{equation}
which is a column-row permuted version of some form of tensor product of single qubit gates and it is easy to see that this matrix is unitary.
This unitary can be applied directly to an ancilla with the marked state defined as $\ket{0}\otimes \ket{\mathbf{v_1}}$.
This generates a quantum state in the following form:
\begin{equation}
\braket{\mathbf{v_0}}{\mathbf{v_1}}\left(
\begin{matrix}
        \ket{\mathbf{v_0}}\\
        \ket{v_1}-\ket{\mathbf{v_0}}
\end{matrix}\right).
\end{equation}
Then in the amplitude amplification, one can  increase the probability by using the known undesired part \ket{\mathbf{v_0}} since the amplitude amplification oscillates: i.e. after the maximum probability, it starts decreasing.
Note that since this matrix requires a circuit decomposition, it is not used in the numerical simulations.
\subsection{Implementation and approximation of $M$ through linear combination of unitaries}
We can implement $M$ by writing as a sum of unitary matrices:
\begin{equation}
\label{Eq:Mlcu}
    M = -\frac{1}{2} X + \frac{1}{2} I.
\end{equation}
Since there are only two terms we can implement this by using only one ancilla qubit and a gate which implements the coefficients. However, since the square root of the coefficients are $1/\sqrt{2}[1, i]$, this implementation also yields to another non-unitary gate and also reduces the complexity gain. 

Similarly $\tilde{M}$ can be implemented by using the following:
\begin{equation}
\label{Eq:Mtildelcu}
       \tilde{M} = \alpha/2 I + \alpha/2 Z +  X -1/2I+1/2Z.
\end{equation}
 This can be implemented by using two ancilla qubits. 
 
To overcome the implementation difficulties, we can also increase the number of terms and the size of the ancilla register. For instance we can introduce $Z$ or $H$ gates to the equation:
\begin{equation}
       M = -\frac{1}{2} X + \frac{1}{2} I +  \beta Z -\beta Z+\eta H -\eta H,
\end{equation}
where $\beta$ and $\eta$ are  coefficients that can be adjusted to make the implementation easier.

One can also map the coefficients to parameterized quantum gates. In this way case the searching problem would become an optimization problem where classical methods can be employed to optimize the parameters.

To get a unitary quantum state result, we can also consider Eq.\eqref{Eq:u1} again. 
It gives the final state. More importantly it maximizes the probability of the solution in this state. Similarly to the above, in this case the coefficients are $[1, -\alpha]$ which may also lead to another non-unitary gate implementations and require additional terms and qubits.

\begin{figure}
\begin{center}
        \Qcircuit @C=1em @R=.7em {
\ket{0}&&\qw & \gate{\tilde{H}} & \qw & \qw& \ctrl{1}&\gate{\tilde{H}}&\qw \\
\ket{0}&&\qw & \gate{H} & \ctrl{1} & \qw& \gate{X}&\qw&\qw \\
\ket{\mathbf{0}}&&/\qw & \gate{H} & \multigate{1}{U_{mark}} & \qw&\qw& \qw&\qw \\
\ket{\mathbf{0}}&&/\qw & \gate{H}& \ghost{U_{mark}} & \qw& \qw & \qw&\qw
}
\end{center}
    \caption{Search algorithm circuit. $M$ is implemented as a sum of unitary gates. $\tilde{H}$ implements coefficients. However, it may require further processing or additional qubits. }
    \label{fig:circuitUnitaryM}
\end{figure}
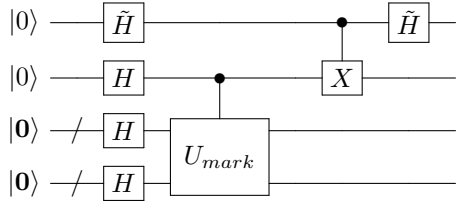

In Fig.\ref{fig:circuitUnitaryM}, we give the base form for the circuit by assuming only two terms are used. Since we only used one Hadamard-like gate, $\tilde{H}$, on the ancilla, the success probability becomes scaled by 1/2.
However, as explained, this operator is also non unitary and therefore more additional qubits may be needed. This would decrease the success probability exponentially in the number of qubits used in the ancilla.

\subsection{Applying the amplitude amplification  to the first qubit}
From Fig.\ref{fig:simulationofC}, we can see that for the small values of $a$, the probability of the first qubit is higher. However, in this cases, the probability of the marked element becomes exponentially small.
On the other hand, if we choose larger $a$ values that are close to 1, such as $(N-2)/N$ or $(n-1)/n$, as it can be seen in the figure, the probability of the marked state when the fist qubit is \ket{0} is maximized.

The small \ket{0} probability of the first qubit can be amplified by using the  amplitude amplification algorithm \cite{brassard2002quantum}. 
We only apply one step of the  amplification by first marking \ket{0} state on the first qubit (i.e. marking the states where the first qubit is \ket{0} in all states). 
We will call this operator ${U_s}$:
\begin{equation}
    U_s = 2\ket{0}\bra{0} - I.
\end{equation}
The global operator becomes:
\begin{equation}
    \mathfrak{U}_s = U_s \otimes I^{\otimes n+k}.
\end{equation}
Note that this operator can be implemented by using a single $Z$ gate. Therefore it does not change the constant complexity of the circuit.

\begin{figure*}[t!]
    \centering
    \begin{subfigure}[t]{0.45\columnwidth}
        \centering
        \includegraphics[width=1\columnwidth]{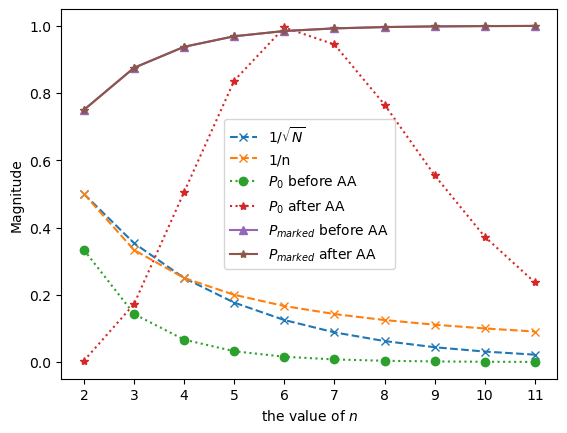}
        \caption{$a = \frac{N-2}{N}$.}
    \end{subfigure}
    \begin{subfigure}[t]{0.45\columnwidth}
        \centering
        \includegraphics[width=1\columnwidth]{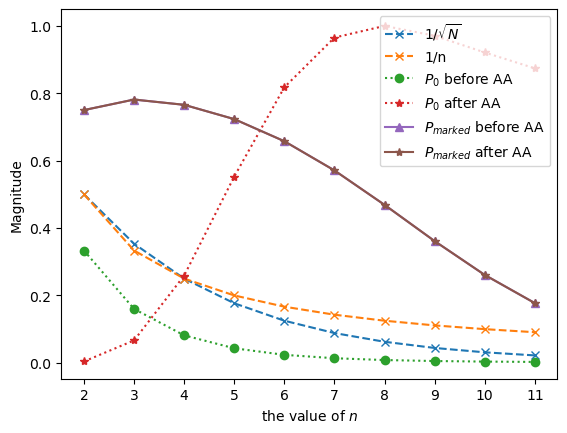}
        \caption{$a = \frac{n-1}{n}$ }
    \end{subfigure}\\
        \begin{subfigure}[t]{0.45\columnwidth}
        \centering
        \includegraphics[width=1\columnwidth]{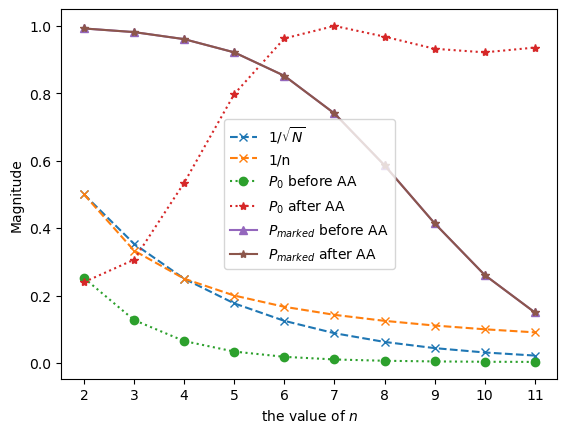}
        \caption{$a = 0.9$.}
    \end{subfigure}
    \begin{subfigure}[t]{0.45\columnwidth}
        \centering
        \includegraphics[width=1\columnwidth]{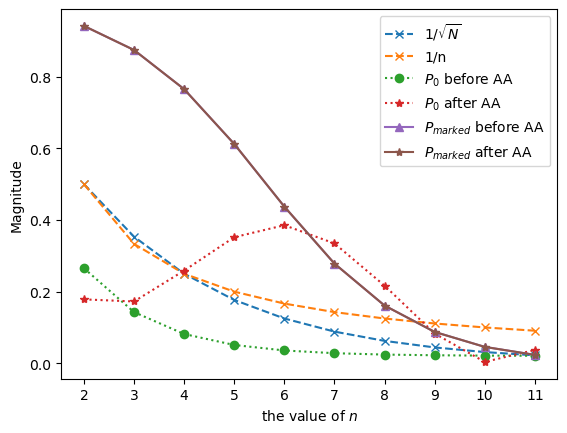}
        \caption{$a = 0.75$. }
    \end{subfigure}
       \caption{\label{fig:simulationofCwithAAn}
       For different $a$ values in matrix $M$ simulation of the circuit with the unitary matrix $\mathfrak{C}$ and \textbf{$n$-step} of the  amplitude amplification.}
    \end{figure*}

The amplification operator is formed by the final output of the whole circuit:
\begin{equation}
    \ket{\psi_{final}} = U_\mathfrak{C} \ket{0},
\end{equation}
where $U_\mathfrak{C}$ is the whole circuit drawn in Fig.\ref{fig:circuitUnitaryC}.  
Then we defined the amplification (the Grover diffusion operator) as:
\begin{equation}
\begin{split}
      U_{aa} & = I - 2\ket{\psi_{final}}\bra{\psi_{final}}.
\end{split}
\end{equation}
This operator does not change the marked elements normalized probabilities. It only changes the probability of the first qubit.
Here, it is important to note that since we are trying to change the state of a single qubit, it is easier to amplify. 
This can be seen in Fig.\ref{fig:simulationofCwithAAn} where we apply the amplification $n$ times, i.e. the number of qubit times.
This can be compared to the plots in Fig.\ref{fig:simulationofC} where we apply only a single amplitude amplification.

From these two figures we can see that, if the initial probability of the marked state in the part where the first qubit is \ket{0} is large, one can simply obtain solution : first by applying amplitude amplification $O(n)$ times to the first qubit and then measuring the first qubit to see this amplification is succeeded.

\begin{figure*}[t!]
    \centering
    \begin{subfigure}[t]{0.45\columnwidth}
        \centering
        \includegraphics[width=1\columnwidth]{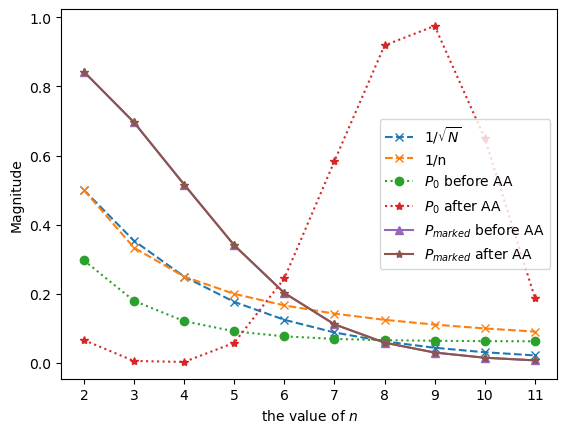}
        \caption{$a = 0.6.$}
    \end{subfigure}
    \begin{subfigure}[t]{0.45\columnwidth}
        \centering
        \includegraphics[width=1\columnwidth]{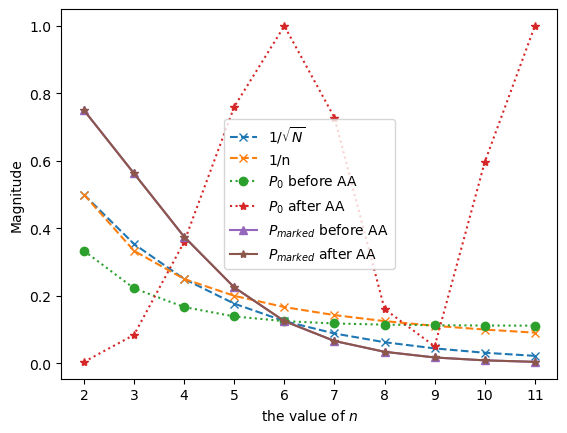}
        \caption{$a = 0.5$ }
    \end{subfigure}
       \caption{\label{fig:simulationofCwithAAnlowa}
       For small $a$ values in matrix $M$, the simulation of the circuit with the unitary matrix $\mathfrak{C}$ and \textbf{$n$-step} of the  amplitude amplification.}
\end{figure*}

If the probability of the marked state in \ket{0} part is not high, that means all $N$ states have almost equal amount of amplitudes. In this case, the amplitude of \ket{0} on the first register can be easily amplified, however, one cannot measure the marked state in the collapsed quantum state.
This is shown in Fig.\ref{fig:simulationofCwithAAnlowa}, it can be seen that steps of the amplification changes the probability dramatically and \ket{0} probability oscillates. However, as it is stated before, even if we measure \ket{0} on the first register, the probability of measuring the marked element is too small.

\section{Searching multiple database simultaneously}
Assume that we have a quantum state with different selected elements on different parts of $K$ number of quantum states that may or may not be different. With the abuse of notation, we will represent each database $i$ by the vector $\mathbf{u_i}$ and the marked version of this vector by $\mathbf{v_i}$. Then the whole quantum state is defined as:
\begin{equation}
   \ket{\psi_1} = \frac{1}{\sqrt{2K}}\left( \begin{matrix}
        \mathbf{u_0}\\
        \mathbf{v_0}\\
        \mathbf{u_1}\\
        \mathbf{v_1}\\
        \vdots\\
        \mathbf{u_K}\\
        \mathbf{v_K}\\
    \end{matrix}
    \right),
\end{equation}
where each $\mathbf{v_i}$ has a marked(selected) element.
Since we have only one element differs in signs, we can easily state that $\|\mathbf{v_0 - v_i}\| =  \frac{2}{\sqrt{N}}$, which is because only one element (the marked) is non-zero in the resulting vector.

To compose $a\mathbf{v_0}-\mathbf{v_i}$ simultaneously for all the choices of $i=1, \dots i=K$; we will use the following block diagonal matrix:
 \begin{equation}
\label{eq:blockdiagonalM}
C  =    \left(\begin{matrix}
     M &&&\\
     &M&&\\
     &&\ddots&\\
     &&&M\\
    \end{matrix}\right)_{2K} \otimes I^{\otimes n}.
\end{equation}
Where I is an $2\times 2$ identity matrix. 
Application of this matrix to the previous vector \ket{\psi_0} generates the following state:
\begin{equation}
        \ket{\psi_2} = \frac{1}{\sqrt{2K}}\left( \begin{matrix}
        a\mathbf{u_0}-\mathbf{v_0}\\
         -\mathbf{u_0}+a\mathbf{v_0}\\
         a\mathbf{u_1}-\mathbf{v_1}\\
         -\mathbf{u_1}+a\mathbf{v_1}\\
        \vdots\\
        a\mathbf{u_k}-\mathbf{v_k}\\
        -\mathbf{u_k}+a\mathbf{v_k}\\
    \end{matrix}\right).
\end{equation}
This vector has $2N$ elements which has higher amplitudes than the rest. In exact terms we can represent the amplitudes of the marked and the unmarked elements by the following $\mu$ and $\nu$ values, respectively:
\begin{equation}
    \mu = \pm \frac{a+1}{\sqrt{2KN}} \text{ and } \nu = \pm \frac{a-1}{\sqrt{2KN}}.
\end{equation}
For each marked element, we have two vector elements with the amplitude $\mu$.

Now, let us first assume $a$ is chosen sufficiently close to 1. And then assume we measure the first $k = log(K)$ qubits and obtain a value $j$ with $0 \leq j\leq K $. Then the state on the remaining $n+1$ qubits collapses on the following state:
\begin{equation}
 \frac{1}{\sqrt{2}}\left( \begin{matrix}
a\mathbf{v_0}-\mathbf{v_j}\\
-\mathbf{v_0}+a\mathbf{v_j}\\
    \end{matrix}
    \right). 
\end{equation}
Since $a$ is close to $1$, a measurement on the last $n$ qubits of this $n+1$ qubit state gives us the marked state with high probability. 

\textbf{In this non-unitary setting}: For $a\approx 1$, that probability of the marked state becomes $P_{marked} \approx 1$. Because, after the measurement, the state is normalized and the only elements that are not close to zero are the marked two elements.

\textbf{In the unitary setting}: The implementation of $C$ will follow the similar ideas discussed in the previous section and thus, we will have an extra qubit which will determine if we obtain above simulation or some another state. The probability of the success for this process and how it can be amplified has been discussed in the previous section.
\subsection{Forming a state of marked elements}
Here note that one can also combine all the different state into a vector of marked state by using the matrix $C$ defined in Eq.\eqref{eq:JotimesM} by using the matrix $J$ defined in Eq.\eqref{eq:matrixJ}.

\section{Complexity analysis and conclusion}
In its base form, the complexity of the circuit is bounded by the complexity of $U_{mark}$. However, as in the original Grover search algorithm, it requires repetitions to either to increase the probability of the first qubit that implements the non-unitary dynamic or the probability of the marked state within the collapsed state.
As shown in the previous sections, these repetitions can be done through the amplitude amplification of \ket{0} state of the first register, which requires choosing $a$ sufficiently close to 1.

In terms of qubits, the algorithm requires one additional qubit in comparison to the original algorithm. However, as it can be seen in Fig.\ref{fig:simulationofC} and Fig.\ref{fig:simulationofCwithAAn} that represents the simulation with different $a$ values, for some cases the approach may be used to obtain directly the solution without or in $O(n)$ repetitions. 
As also discussed in  \cite{lloyd2000ultimate}, there are limits to the computations and so NP-complete problems cannot be solved exponentially faster on quantum computers unless $P = NP$ \cite{aaronson2005guest}.

Therefore, although  the approach gives a concrete implementation alternative to standard circuit, the probabilities will be much lower when $n$ grows as shown in the figures. Therefore in those cases it may not be feasible to use the circuit.

\section{Data Availability}

The simulation code can be accessed from \url{https://github.com/adaskin/non-unitary-search.git} which is used to generate figures.

\section{Funding}
This project is not funded by any funding agency.

\appendix

\section{Grouping Set Elements}
Consider a set containing two groups of elements, red and blue, where each element has an associated probability $p_i$. Our goal is to separate these elements into two disjoint sets:
\begin{center}
\includegraphics[width=4in]{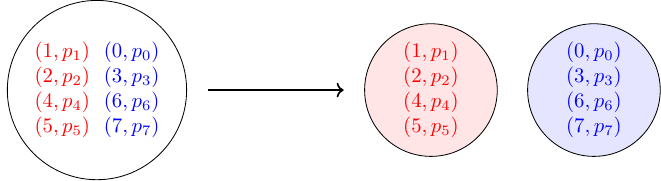}
\end{center}
This problem can be formulated for quantum computers by first constructing two quantum states: $\vec{u}{init}$ representing the elements with their probabilities, and $\vec{u}{marked}$ which uses sign differences in the amplitudes to distinguish between element types:
\begin{equation}
    \vec{u}_{init} = \begin{bmatrix}
        p_0\\ p_1\\ p_2\\ p_4\\ p_5\\ p_6\\ p_7\\ 
    \end{bmatrix} \text{ and } \  \vec{u}_{marked} = \begin{bmatrix}
        {\color{blue}-p_0}\\ {\color{red}p_1}\\ {\color{red}p_2}\\ {\color{blue}-p_3}\\ 
        {\color{red}p_4}\\ {\color{red}p_5}\\ {\color{blue}-p_6}\\ {\color{blue}-p_7}\\ 
    \end{bmatrix}.
\end{equation}
The solution to this problem is then represented by the following two quantum states:
\begin{equation}
{\color{blue}
    \vec{s}_{marked} = \begin{bmatrix}
        p_0\\ 0\\ 0\\ p_3\\ 0\\ 0\\ p_6\\ p_7\\ 
    \end{bmatrix} } \text{ and }    {\color{red}\vec{s}_{unmarked} = \begin{bmatrix}
        0\\ p_1\\ p_2\\ 0\\ p_4\\ p_5\\ 0\\ 0\\ 
    \end{bmatrix}}.
\end{equation}
We construct the quantum state $\vec{u}_{init}$ to encode the probability amplitudes of the elements. For example, given a set, we prepare a quantum state representing the likelihood of each element:
Consider $2^n$-dimensional vectors $\vec{u}{init}$ and $\vec{u}{marked}$ that differ only in the signs of certain elements, marking their group membership: 
For $a \in R$ and $a\geq 0.5$,  if the following operator, $\mathfrak{C}$, is applied to a vector horizontally stacked with $\vec{u}_{init}$ and $\vec{u}_{marked}$ and zero vectors, on the first half of the output we get one group and on the remaining half of the output is the second group elements with higher probabilities:
\begin{equation}
\underbrace{
\frac{1}{a+1}
\begin{bmatrix}
a & -1 & -\sqrt{a}& -\sqrt{a}\\
-1 & a & -\sqrt{a} & -\sqrt{a}\\
\sqrt{a}& \sqrt{a} &a & -1\\
\sqrt{a} & \sqrt{a} & -1 & a
\end{bmatrix}}_{\mathfrak{C}} \otimes I^{\otimes n} \times
\begin{bmatrix}
    \vec{u}_{marked}\\ \vec{u}_{init}\\ \hline \vec{0}\\ \vec{0}
\end{bmatrix}= 
\frac{1}{a+1}
\begin{bmatrix}
   a\vec{u}_{marked}-\vec{u}_{init}\\-\vec{u}_{marked}+au_{init}\\ 
   \hline\sqrt{a}(\vec{u}_{marked}+\vec{u}_{init})
   \\ \sqrt{a}(\vec{u}_{marked}+\vec{u}_{init}) 
\end{bmatrix}.
\end{equation}
When $a$ is close to 1 (e.g. $(n-1)/n$),  this can be used to indicate separate two group of elements by the value of the first ancilla qubit:
\begin{equation}
  \frac{1}{2}
\begin{bmatrix}
   u_{marked}-u_{init}\\-u_{marked}+u_{init}\\ \hline (u_{marked}+u_{init})\\(u_{marked}+u_{init}) 
\end{bmatrix} =
\frac{1}{2}
\begin{bmatrix}
  \color{blue} -s_{marked}\\ \color{blue}s_{marked}\\ \hline \color{red}s_{unmarked} \\\color{red} s_{unmarked} 
\end{bmatrix}
\end{equation}
When measuring the first qubit of this quantum state, outcome \ket{0} yields elements from one group, while outcome \ket{1} yields elements from the second group in the collapsed state.
The probability for both groups can be amplified through the amplitude amplification on the ancilla. 

\textbf{Separating multiple groups.} Similarly, after marking different groups, we can stack them all together to have a quantum state where ancilla indicates different groups in the set.
\begin{equation}
    \begin{bmatrix}
        \mathfrak{C} &&&\\
        &\mathfrak{C}&&\\
        &&\ddots&&\\
        &&&\mathfrak{C}\\
    \end{bmatrix} \otimes I^{\otimes n} \times 
    \begin{bmatrix}
    \vec{u}^{(1)}_{marked}\\ \vec{u}_{init}\\ \vec{0}\\ \vec{0}\\
    \hline \vdots\\
    \hline\vec{u}^{(k)}_{marked}\\ \vec{u}_{init}\\ \vec{0}\\ \vec{0}\\
\end{bmatrix} = 
\begin{bmatrix}
  \color{blue} -s^{(1)}_{marked}\\ \color{blue}s^{(1)}_{marked}\\ \color{red}s^{(1)}_{unmarked} \\\color{red} s^{(1)}_{unmarked}
  \\ \hline \vdots \\ \hline 
  \color{blue} -s^{(k)}_{marked}\\ \color{blue}s^{(k)}_{marked}\\  \color{red}s^{(k)}_{unmarked} \\\color{red} s^{(k)}_{unmarked}
\end{bmatrix} \xrightarrow{SWAP}
\begin{bmatrix}
  \color{blue} -s^{(1)}_{marked}\\ \color{blue}s^{(1)}_{marked}\\ \vdots\\  
  \color{blue} -s^{(k)}_{marked}\\ \color{blue}s^{(k)}_{marked}\\ \hline \color{red}s^{(1)}_{unmarked} \\\color{red} s^{(1)}_{unmarked}
\\ \vdots \\
\color{red}s^{(k)}_{unmarked} \\\color{red} s^{(k)}_{unmarked}
\end{bmatrix}
\end{equation}
This quantum state represents the sets in the upper half and their complements (the unmarked elements) in the second half.
Note that instead of the same  $\vec{u}_{init}$, one can use different $\vec{u}_{init}$ in a similar fashion to operate on different sets.

\begin{figure}[ht]
    \centering
    \includegraphics[width=0.5\linewidth]{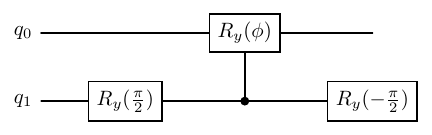}
    \caption{Explicit circuit implementation of $\mathfrak{C}$, where $R_y$ denotes the standard Y-rotation gate. The parameter $\phi$ depends on $a$: when $a=1$, $\phi=\pi$; for other values, $\phi$ is chosen to satisfy $R_y(\phi) = \begin{bmatrix} \frac{a-1}{a+1} & -\frac{2\sqrt{a}}{a+1} \\ \frac{2\sqrt{a}}{a+1} & \frac{a-1}{a+1} \end{bmatrix}$. }
    \label{fig:circuit_c}
\end{figure}

\subsection{Quantum circuit implementation of $\mathfrak{C}$}
\begin{figure}
    \centering
    \includegraphics[width=0.75\linewidth]{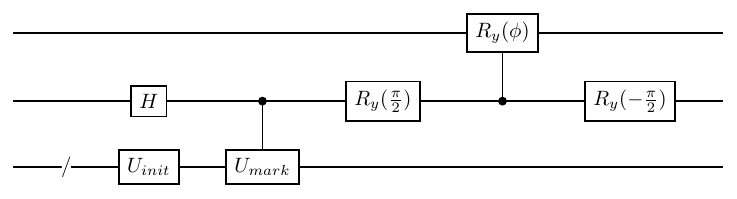}
    \caption{The whole direct search circuit without amplitude amplification. The amplitude amplification can be applied by marking the desired set on the ancilla.}
    \label{fig:whole-circuit}
\end{figure}

We begin by defining the cosine ($C$) and sine ($S$) matrices:
\begin{equation}
    C = \frac{1}{a+1}\begin{bmatrix}
        a&-1\\
        -1&a
    \end{bmatrix} \text{ and } 
    S = \frac{\sqrt{a}}{a+1}\begin{bmatrix}
        1&1\\1&1 
    \end{bmatrix}.
\end{equation}
The unitary operator $\mathfrak{C}$ can then be expressed in cosine-sine form as:
\begin{equation}
\mathfrak{C} = \frac{1}{a+1}
\begin{bmatrix}
a & -1 & -\sqrt{a}& -\sqrt{a}\\
-1 & a & -\sqrt{a} & -\sqrt{a}\\
\sqrt{a}& \sqrt{a} &a & -1\\
\sqrt{a} & \sqrt{a} & -1 & a
\end{bmatrix} =  
\begin{bmatrix}
C & -S \\
S & C
\end{bmatrix} \text{ with } C^2 + S^2 = I. 
\end{equation}
Here, $C^2 + S^2 = I$ ensures unitarity.
The eigenspaces for both matrices are characterized by the Hadamard like matrices with the following decompositions:
\begin{equation}
    S = \left( \begin{matrix}
       1& 1\\
       -1 &1
    \end{matrix}\right)\left( \begin{matrix}
       0& 0\\
       0 &\frac{\sqrt{a}}{a+1}
    \end{matrix}\right)
    \left( \begin{matrix}
       1& -1\\
       1 &1
    \end{matrix}\right) =  \frac{1}{a+1} \left( \begin{matrix}
       \sqrt{a}& \sqrt{a}\\
       \sqrt{a} &\sqrt{a}
    \end{matrix}\right). 
\end{equation}
\begin{equation}
\begin{split}
    C =  & \frac{1}{2(a+1)}\left( \begin{matrix}
       1& 1\\
       -1 &1
    \end{matrix}\right)\left( \begin{matrix}
       a+1& 0\\
       0 &a-1
    \end{matrix}\right)
    \left( \begin{matrix}
       1& -1\\
       1 &1
    \end{matrix}\right) \\ =  & \frac{1}{2(a+1)}\left( \begin{matrix}
       a+1& a-1\\
       -(a+1) &a-1
    \end{matrix}\right)
    \left( \begin{matrix}
       1& -1\\
       1 &1
    \end{matrix}\right)\\ = &\frac{1}{a+1}\left( \begin{matrix}
       a& -1\\
       -1 &a
    \end{matrix}\right). 
\end{split}
\end{equation}
More specifically, for the quantum gate \(V = \frac{1}{\sqrt{2}} \begin{bmatrix} 1 & 1 \\ -1 & 1 \end{bmatrix}\); the eigendecompositions can be finalized as:
\begin{equation}
C = V D_C V^T, \quad S = V D_S V^T,
\end{equation}
with 
\begin{equation}
D_C = \begin{bmatrix} 1 & 0 \\ 0 & \frac{a-1}{a+1} \end{bmatrix}, \quad D_S = \begin{bmatrix} 0 & 0 \\ 0 & \frac{2\sqrt{a}}{a+1} \end{bmatrix}.
\end{equation}
Therefore, \(\mathfrak{C}\) can be factorized into:
\begin{equation}
\mathfrak{C} = (I_2 \otimes V) \cdot \mathcal{D} \cdot (I_2 \otimes V^T),
\end{equation}
where \(\mathcal{D}\) is a block-diagonal matrix:
\begin{equation}
\mathcal{D} = \begin{bmatrix} D_C & -D_S \\ D_S & D_C \end{bmatrix}.
\end{equation}
We leverage the cosine-sine decomposition and eigendecompositions of $C$ and $S$ to construct a quantum circuit for the operator \(\mathfrak{C}\). 
In the basis defined by \(V^T\), \(\mathcal{D}\) simplifies to the following:
\begin{itemize}
\item When the second qubit is \(|1\rangle\), to the first qubit, apply the following gate: \begin{equation}R = \begin{bmatrix} \frac{a-1}{a+1} & -\frac{2\sqrt{a}}{a+1} \\ \frac{2\sqrt{a}}{a+1} & \frac{a-1}{a+1} \end{bmatrix}.\end{equation} 
\item When the second qubit (in the eigenbasis) is \(|0\rangle\), do nothing or apply \(I\) to the first qubit.
\end{itemize}
Therefore, the whole quantum circuit is defined as the sequence requiring only 3 operations:
\begin{enumerate}
    \item Applying $V^T = R_y(90^\circ) = R_y(\pi/2)$ gate to the second qubit in order to change the basis. 
    \item Applying  the following rotation gate:
\begin{equation}R = R_y(\phi) = \begin{bmatrix} \frac{a-1}{a+1} & -\frac{2\sqrt{a}}{a+1} \\ \frac{2\sqrt{a}}{a+1} & \frac{a-1}{a+1} \end{bmatrix}\quad \text{where} \quad \cos(\phi/2) = \frac{a-1}{a+1}, \quad \sin(\phi/2) = \frac{2\sqrt{a}}{a+1}.
\end{equation}
\begin{itemize}
    \item 
For \(a = 1\)(\textbf{should choose a close value to 1, e.g. $(n-1)/n$, since 1 makes  the eigenvalue in $D_C$ zero}):
\begin{equation}
\cos(\phi/2) = \frac{1-1}{1+1} = 0, \quad \sin(\phi/2) = \frac{2\sqrt{1}}{1+1} = 1 \implies \phi/2 = \pi/2, \quad \phi = \pi.
\end{equation}
Thus, \(R_y(\pi) = \begin{bmatrix} 0 & -1 \\ 1 & 0 \end{bmatrix}\) (a Pauli-Y rotation up to phase).

\end{itemize}
\item Applying $V = R_y(-90^\circ) = R_y(-\pi/2)$ gate to the second qubit reverts the basis back.
\end{enumerate}.
By combining the circuit for $\mathfrak{C}$ with the initial state preparation and marking operation, we can define the whole framework as in Fig.~\ref{fig:whole-circuit}.

\subsection{Complexity Analysis}
The whole circuit given in Fig.\ref{fig:whole-circuit} requires $O(n)$ quantum operations if the marking circuit $U_{mark}$ (which is generally implemented through a controlled-Z gate along with a function output) and $U_{init}$ the state preparation operation (Hadamard gates for an equal superposition state) are in $O(n)$. The operator $\mathfrak{C}$ requires only 3 gates per qubit pair.

At the end, the probabilities can be amplified by using amplitude amplification algorithm. If the amplification is only for a single element, this would take $O(\sqrt{2^n})$.

\bibliographystyle{unsrtnat}
\bibliography{paper}

\end{document}